\documentclass[fleqn,10pt
]{wlpeerj}
\usepackage{enumitem}
\usepackage{verbatim}
\usepackage{subcaption}
\usepackage{nameref}
\title{LazyFox: Fast and parallelized overlapping community detection in large graphs}

\author[1]{Tim Garrels}
\author[1]{Athar Khodabakhsh}
\author[1,2]{Bernhard Y.\ Renard}
\author[1]{Katharina Baum}
\affil[1]{Hasso Plattner Institute, Digital Engineering Faculty, University of Potsdam, Prof.-Dr.-Helmert Str. 2-3, 14482 Potsdam, Germany}
\affil[2]{Department of Mathematics and Computer Science, Free University Berlin, Takustraße 9, 14195 Berlin, Germany}
\corrauthor[]{Katharina Baum}{katharina.baum@hpi.de}
\usepackage{amsmath}
\usepackage{amssymb}
\usepackage{caption}
\usepackage{graphicx}
\usepackage{hyperref}
\usepackage{listings}
\usepackage{xspace}

\usepackage{algorithm} 
\usepackage{algpseudocode} 

\newcommand{\fox}{\textsc{Fox}\xspace}
\newcommand{\lazyfox}{\textsc{LazyFox}\xspace}

\newcommand{\wcchat}{\ensuremath{\widehat{WCC}}\xspace}

\newcommand{\lyu}{Lyu and colleagues\xspace}

\newcommand{\snapdownloaded}[1]{\footnote{\url{https://snap.stanford.edu/data/#1.txt.gz}, retrieved \date{15.09.2021}}}

\begin{abstract}
The detection of communities in graph datasets provides insight about a graph's underlying structure and is an important tool for various domains such as social sciences, marketing, traffic forecast, and drug discovery. While most existing algorithms provide fast approaches for community detection, their results usually contain strictly separated communities. However, most datasets would semantically  allow for or even require overlapping communities that can only be determined at much higher computational cost. We build on an efficient algorithm, \fox, that detects such overlapping communities. \fox measures the closeness of a node to a community by approximating the count of triangles which that node forms with that community. We propose \lazyfox, a multi-threaded version of the \fox algorithm, which provides even faster detection without an impact on community quality. This allows for the analyses of significantly larger and more complex datasets. \lazyfox enables overlapping community detection on complex graph datasets with millions of nodes and billions of edges in days instead of weeks. As part of this work, \lazyfox's implementation was published and is available as a tool under an MIT licence at \url{https://github.com/TimGarrels/LazyFox}.
\end{abstract}

\begin{document}

\flushbottom
\maketitle
\thispagestyle{empty}

\section*{Introduction \& Related Work}
Graphs, also called networks, are present in many fields as they capture the interaction of entities as edges between nodes representing those entities. Communities (or clusters, or modules) are considered to be important network structures as they describe functionally similar groups within networks. The identification of such groups is relevant in various domains such as biology \citep{Barabasi2004NetworkBiology, Regan2002HierarchicalMetabolic, Boccaletti2006ComplexNetworks, Guimera2005FunctMetabolic, Ahn2010MultiscaleComplex}, medicine \citep{Barabasi2011NetworkMedicine, Gavin2006YeastModularity} and technical infrastructure \citep{Regan2003HierachyComplexNetworks, Guimera2005AirTransportNetworks}.
In social networks, communities can identify common interest groups or friend circles \citep{Mcauley2014SocialCircles}. In drug research, new, targetable proteins can be discovered by clustering proteins into functionality groups \citep{Ma2019PPIdrugsdiscovery}. Similar products of an online shop can be grouped together to derive product recommendations \citep{Basuchowdhuri2014ProductRecommendation}.

While community itself is a term whose precise definition is highly context dependent \citep{Kelley2012SocialCommunities}, various algorithms have been developed to detect such structures \citep{Fortunato2010}. In this work, we focus on communities of bond. These are communities where the membership of a community is based on the relation to other members -- their connectivity in the network -- rather than an affinity to the identity of that group as a whole. This naturally leads to more interconnections between group members \citep{Ren2007}. Algorithms detecting such communities typically exploit this property by optimizing communities so that nodes of one group are more densely connected with each other than with nodes outside of that group.

\subsection*{Disjoint Community Detection}
Disjoint communities capture structural node clusters within a network where each node belongs to exactly one community. While algorithms based on label propagation \citep{Raghavan2007LabelPropagation}, stochastic blockmodels \citep{hofman2008bayesModularity, Wang1987StochasticBF} and game theory \citep{Bu2017GLEAM} for disjoint community detection exist, the most common approach is to define a metric that measures desired structural properties of a community \citep{Lancichinetti2011OSLOM, Yang2011CommunityMetricEval, Newman2006Modularity, wcc}. By choosing communities that increase the metric over communities that decrease it, the communities are optimized to display that structural property. This way a clear community definition (groups of nodes that result in the highest value of the metric) is created and detected communities are to some degree controllable in their structure.

\subsection*{Overlapping Community Detection}
Most approaches identify communities in such a way that each node can belong to only one community. While disjoint communities yield valuable information, most real world datasets contain functional groups that are overlapping \citep{Reid2011PartitioningBC, Lancichinetti2009DetectingTO}. People can belong to multiple social circles, proteins can have various biological functions or be target of multiple drugs, and products rarely fall into exactly one category. 
Therefore, traditional community detection algorithms have been adapted and new algorithms have been proposed to detect overlapping communities \citep{Fortunato2010}. However, overlapping community detection is a computationally more expensive problem than regular (disjoint) community detection. Long runtimes of days or even weeks make community detection unfeasible in large datasets, which is especially true for overlapping community detection \citep{Lancichinetti2009ComparativeComDetection, Danon2005ComparingCommunityI}. This limits the usability of most algorithms to smaller sized networks with edge counts in the lower millions.

Algorithms which detect overlapping communities can use other approaches than disjoint community detection algorithms such as clique expansion \citep{Lee2010DetectingHO, palla2005cfinder}, matrix factorization \citep{Yang2013OverlappingCD, Psorakis2011OverlappingCD} and link clustering \citep{Shi2013ALC, Ahn2010MultiscaleComplex, evans2009LinkPartitions}. Some approaches also propose post-processing steps to create overlapping communities from pre-existing disjoint communities \citep{Chakraborty2015LeveragingDC}.

Moreover, approaches that have been proposed to detect disjoint communities can also be adjusted to directly yield overlapping communities. Label propagation based algorithms can allow nodes to hold multiple labels at the same time, thereby creating overlapping communities \citep{Xie2012OverlapLabelProp, Gregory2009FindingOC}. Stochastic blockmodels can support mixed memberships \citep{airoldi2007mixed, Gopalan14534}. And if an algorithm is based on the optimization of a metric measuring structural properties, allowing nodes to join multiple communities during the optimization can also change the result from disjoint to overlapping communities \citep{Lancichinetti2009DetectingTO}.

\subsection*{Enabling Large Scale Community Detection}
Large scale datasets pose a challenge for many community detection algorithms as their runtime increases drastically \citep{Lancichinetti2009ComparativeComDetection, Danon2005ComparingCommunityI}. Various approaches can be used to improve runtime and enable the use of community detection algorithms on large datasets. Parallelization can leverage modern multi-core CPUs, and algorithms with independent computation steps can profit from parallelization directly \citep{Liu2018SpeedingUB}. Other algorithms cannot be parallelized without logical changes, however, small alterations often do not influence the results substantially \citep{PratPerezSCD}. Such parallelized approaches can also be distributed to a multi-machine computing cluster, enabling the use of even more computational resources \citep{Saltz2015DistributedCD}.
Finally, algorithms relying on metric optimization can also be improved in runtime by replacing the metric with an estimator. This reduces computational effort and speeds up the community detection.

\paragraph{Weighted Community Clustering Score}
The Weighted Community Clustering ($WCC$) score \citep{wcc} is a metric that has been successfully used to detect communities \citep{Saltz2015DistributedCD, Song2015, PratPerezSCD}. It measures structural closeness by counting the number of triangles a node forms with a community to determine the membership of that node to that community. Thereby, it is closely related to the general concept of clustering coefficients. In \hyperref[sec:methods]{Methods} we explain this metric in more detail. Community detection algorithms based on this metric optimize communities to maximize the global $WCC$ score by allowing nodes to individually join or leave communities. Such algorithms can yield disjoint or overlapping communities depending on whether nodes are allowed to join multiple communities.

\paragraph{Efficient $WCC$-based large scale overlapping community detection}
$WCC$-based overlapping community detection approaches also suffer from computational complexity issues when dealing with large scale datasets. Thus, parallelized \citep{PratPerezSCD} and distributed \citep{Saltz2015DistributedCD} $WCC$ versions have been proposed that allow multiple nodes at the same time to decide for changes to the community partition, so those decisions can be computed on separate threads or machines. Furthermore, computationally less expensive estimations of the $WCC$ score have been suggested \citep{PratPerezSCD, fox} to further improve runtime. Thereby, \lyu have found that approximating triangle counts with their metric, \wcchat, in an algorithm they call \fox yields very similar results as calculating the exact count. While their proposed estimation allows \fox to compute overlapping communities on large networks with millions of edges, the implementation of \fox was not published making it hard to reproduce the results or to use them for further applications.

\paragraph{\lazyfox - Parallelized Large Scale Overlapping Community Detection}
We here propose our tool for overlapping community detection, \lazyfox, that contains an implementation of the \fox algorithm. In addition, we extend the work of \lyu and combine it with ideas from \citep{PratPerezSCD} by introducing parallelism to leverage the advantage of modern multiprocessor systems to even further accelerate the algorithm. 

We show that \lazyfox produces extremely similar results to \fox in a fraction of \fox's runtime, making it more efficient to work with. This also enables the usage of the approach on large scale graphs like the social media network Friendster with millions of nodes and billions of edges \citep{snapnets}. We analyse the impact of hyperparameters of the \fox and \lazyfox algorithm on real-world examples. The C++ implementation of \lazyfox (which also includes a \fox implementation) is published as open-source under MIT licence.

In the following sections we introduce the datasets, the algorithmic details of \fox and \lazyfox and the methods used to evaluate them. We describe the impact of our \lazyfox contribution on the runtime and community quality of three datasets. We include the analysis of a fourth dataset to illustrate the ability of \lazyfox to handle large scale datasets in contrast to \fox. Finally, we summarize and discuss our results.


\section*{Methods}
\label{sec:methods}
We introduce \lazyfox, with its required input data and preprocessing, the employed metric \wcchat, and the performed steps in the algorithm. We describe the parallelization that sets \lazyfox apart from \fox \citep{fox} and allows \lazyfox to scale across multiple CPU cores, and finally the performance measures employed here.

\subsection*{Input Data}
Our algorithm \lazyfox takes the edge list of an undirected, unweighted graph $G(V,E)$ as input, $V$ denoting the nodes and $E$ the edges of the graph. We assume this graph to be connected for the theoretical discussions, however, \lazyfox will still work on a graph with multiple connected components.

\subsection*{Preprocessing}
\label{methods_preprocessing}
\lazyfox performs the following preprocessing steps on the input edge list of $G$:

\paragraph{Remove Multi-Edges} 
In graphs a pair of nodes can be connected via multiple edges. Since our research on \fox and \lazyfox works on ``by-bond" communities, we regard multiple connections between two entities as one. Therefore, we remove such duplicated edges, turning any input multi-graph into a simple graph.

\paragraph{Dense Node Labels}
The node identifiers (IDs) in edge lists are not necessarily consecutive due to preprocessing (this is the case in most datasets introduced in the 'Datasets' section). We apply a shift to the node IDs to restore their consecutive nature. This allows both \fox and \lazyfox to use the node IDs directly as data-structure indices which speeds up computation.

\paragraph{Node Order}
As \lazyfox computes communities by gradually improving node memberships, there needs to be a well defined order in which the nodes are being processed. We define this order by sorting them by decreasing clustering coefficient (CC):
\[CC(i) = \begin{cases} 
\frac{k_i}{d_i \cdot (d_i - 1) / 2}, & \text{if } d_i > 1,\\
0, & \text{else}
\end{cases}
\]
$k_i$ denotes the number of triangles containing node $i$ (equal to the number of edges between two neighbors of node $i$), while $d_i$ denotes the degree of node $i$. A higher CC indicates that a node is central in its neighborhood, and we process more central nodes first. Ties are resolved by node degrees in decreasing order. This order ensures that nodes with less connectivity and thus less influence can adapt to changes in memberships of more important, connected nodes resulting in a more coherent community structure at the end of one iteration.

\paragraph{Initial Clustering}
To initiate the clustering process \lazyfox computes a greedy, non-overlapping community decomposition using the above node order. A node that is not yet part of a community is assigned to a new community. Then all its not yet assigned neighbors join this community.

This process allows for self-initialization on any given network. \lazyfox derives the initial community count and the initial clustering from the structure of the underlying network. On the other hand, \lazyfox is, just like \fox \citep{fox}, also able to improve upon an existing division into communities by replacing the initial clustering step by inserting those existing divisions. Therefore, known structural properties can be taken into account. This way \lazyfox can be used to generate overlapping community structure from partitions, i.e.\ from non-overlapping community structure. See \cite{fox} for a discussion and examples.

\subsection*{\wcchat -- A Metric to Optimize for}
\label{subsec:wcchat}
\lyu introduced \wcchat as an advanced metric to assess the quality of a partition into communities. This metric forms the core of both the \fox and the \lazyfox algorithm as they use \wcchat to decide on the necessary local optimization steps.
To provide a better understanding of \wcchat we will first describe $WCC$.

\paragraph{WCC}
The weighted clustering coefficient $WCC$ as introduced by \cite{wcc} is a score that rates a community decomposition as the sum of its community ratings. We denote such a rating of a community decomposition $P = \{C_1, \dots, C_k\}$ as \[
WCC(P) = \sum_{i=1}^k WCC(C_i)
\]
This rating of an individual community $C_i$ can be again decomposed into how well the individual nodes of that community fit into the community:
\[ 
WCC(C_i) = \sum_{x \in C_i} WCC(x, C_i)
\]

Such a fit of a node $x$ into its community $C$ is assessed by the ratio of $t(x,C)$ to $t(x,V)$: The number of triangles that the node forms within its community to the number of triangles that it forms within the whole graph.
The assessment also uses $vt(x, C)$, the number of nodes in the community $C$ that form triangles with node $x$. Furthermore, $vt(x, V \setminus C)$, the number of nodes outside of the community $C$ that form triangles with node $x$, influences this assessment:
\[WCC(x, C) = \begin{cases}  
\frac{t(x,C)}{t(x,V)} \cdot \frac{vt(x, V)}{ \mid C \setminus \{x\} \mid + vt(x, V \setminus C)}, & \text{if } t(x,V) > 0,\\
0 & \text{else}
\end{cases} \]

\paragraph{Heuristic Counting of Triangles}
The counting of triangles for the $WCC$ computation is expensive. To accelerate the evaluation of a new community decomposition, the exact count can be replaced by a heuristic. Rather than counting the triangles a node forms with the nodes of a community $C$ and all other nodes, \wcchat \citep{fox} approximates this by using the number of edges instead.

Assuming that the edges inside a community are homogeneously distributed between the nodes, a mean-field approximation can be performed. This delivers the expected triangle count depending on the density $p$ of the community and the degree $deg(x,C)$, which is the count of edges between node $x$ and all nodes within community $C$:

\[\mathbb{E}[t(x,C)] = \binom{deg(x,C)}{2} \cdot p\]
Analogously, one can also define an estimator for the average number of triangles in the whole graph.
It uses the global clustering coefficient $cc$ \citep{Watts1998}, the average of all local clustering coefficients (CC):
\[\mathbb{E}[t(x,V)] = \binom{deg(x,V)}{2} \cdot cc\]

The exact values of  $vt(x, V)$ and $vt(x, V \setminus C)$ are harder to estimate. However, both have an upper bound. If all neighbours of $x$ form a triangle with $x$, the upper bound of $vt(x, V)$ is reached -- it is exactly $deg(x,V)$. The same holds for all neighbors of the vertices $V \setminus C$, so that the upper bound of $vt(x, V \setminus C)$ is $deg(x,V \setminus C)$.

Using $\mathbb{E}[t(x,C)]$, $\mathbb{E}[t(x,V)]$ and these upper bounds, an estimator \wcchat for the weighted clustering coefficient WCC can be derived:
\[\wcchat(x,C) = \begin{cases}
\frac{\mathbb{E}[t(x,C)]}{\mathbb{E}[t(x,V)]}  \frac{deg(x,V)}{\mid C \setminus \{x\} \mid + deg(x, V \setminus C)}, & \text{if } \mathbb{E}[t(x,V)] > 0, \\
0, & \text{else}
\end{cases}\]

This can be then used to define the \wcchat for whole communities $C_i$ and community decompositions $P = \{C_1, \dots, C_k\}$, respectively:
\[
\wcchat(C_i) = \sum_{x \in C_i} \wcchat(x, C_i)
\]
\[
\wcchat(P) = \sum_{i=1}^k \wcchat(C_i)
\]

Both \fox \citep{fox} and \lazyfox use this estimate to compute the optimization steps, as it is faster than the WCC metric.

\subsection*{Node Changes}
Starting with the initial, non-overlapping community decomposition obtained by the preprocessing step, \fox and \lazyfox process the nodes of our graph in multiple iterations. One iteration here is equivalent to computing and applying changes for each node once, described in \autoref{fox_iteration_algorithm} and \autoref{layzfox_iteration_algorithm}. Before deciding on changes, \lazyfox gathers $\wcchat(P)$, the quality rating of the current decomposition $P$.
The \lazyfox algorithm then computes for the node a potential \emph{join}- and a potential \emph{leave}-action.

\paragraph{Joining a Community}
For the current node $x$, \lazyfox finds all communities that $x$ is currently not part of but any of $x$'s neighbors are. Each of these communities $C_k$ is then evaluated by creating a decomposition $P'$ that differs from the current one by adding $x$ to $C_k$ and then computing the,  potentially negative, improvement in $\wcchat(P')$ compared to $\wcchat(P)$. If any of these changes yields a positive improvement, we choose the community with the highest increase in \wcchat as the current \emph{join}-action. This is described in lines two to six in \autoref{fox_iteration_algorithm}.

\paragraph{Leaving a Community}
Similarly \lazyfox checks all communities $C_k$ of the current node $x$ and evaluates the gain in \wcchat if $x$ leaves that community. Therefore, we again form a $P'$ per community by removing $x$ from $C_k$ and gather the improvement in \wcchat compared to the old decomposition. If any of these improvements is positive, we again choose the community with the highest increase in \wcchat as the current \emph{leave}-action. This is described in lines seven to eleven in \autoref{fox_iteration_algorithm}.

After determining the best \emph{join}-action and \emph{leave}-action for a node, these are executed yielding a new decomposition as the new $P$ before calculating change actions for the next node. This is described in line twelve.

\begin{figure}[ht]
\centering
\begin{subfigure}[b]{1\textwidth}
   \includegraphics[width=1\linewidth]{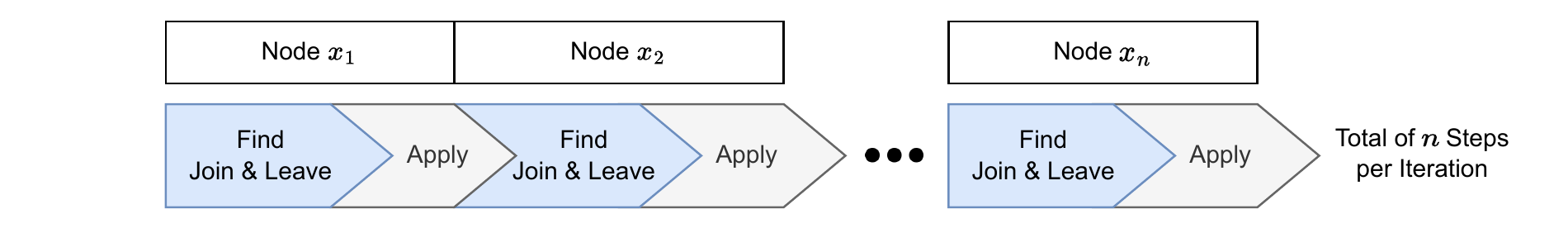}
   \caption{\fox iteration}
   \label{fox_flow}
\end{subfigure}

\begin{subfigure}[b]{1\textwidth}
   \includegraphics[width=1\linewidth]{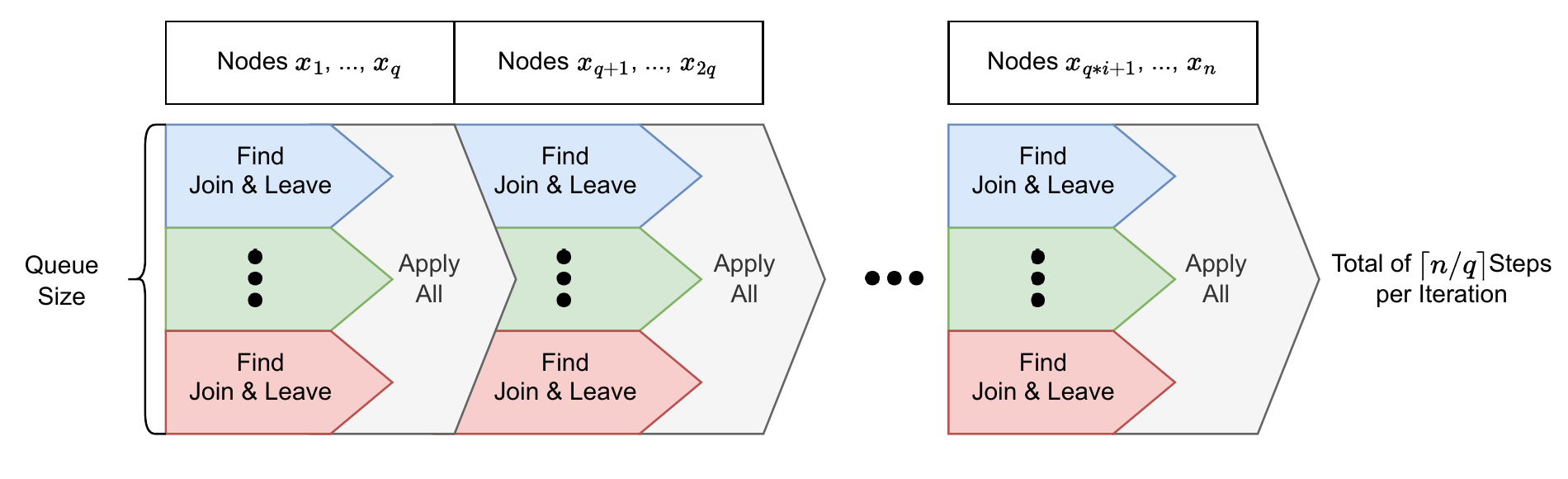}
   \caption{\lazyfox iteration}
   \label{lazyfox_flow}
\end{subfigure}

\caption{Program Control Flows}
\small
Each color denotes a separate thread. While \fox (\autoref{fox_flow}) is computed sequentially on one thread, \lazyfox (\autoref{lazyfox_flow}) can leverage up to queue size many threads that are synchronized by applying their respective node changes.  
Queue size is denoted as $q$. Note that each Find-Apply-Block in \lazyfox contains $q$ nodes, except for the last one, as the total number of nodes of the graph, $n$, might not be a multiple of $q$.
\end{figure}

In \fox, the steps of one iteration are sequential and executed on one single thread, visualized in \autoref{fox_flow}.

\begin{algorithm}
	\caption{\fox \citep{fox}}
    \label{fox_iteration_algorithm}
	\begin{algorithmic}[1]
		\For {node $x$ $\in$ Graph}
			\State bestJoin = undefined
			\For {community $c$ that does not contain $x$ and is a neighbor of $x$}
				\State join = partition if $x$ joined $c$
				\State update bestJoin if the join partition has a higher $\wcchat$ score
			\EndFor
		    \State bestLeave = undefined
			\For {community $c$ that contains $x$}
				\State leave = partition if $x$ left $c$
				\State update bestLeave if the leave partition has a higher $\wcchat$ score
			\EndFor
			\State apply bestLeave and bestJoin to the partitioning
		\EndFor
	\end{algorithmic} 
\end{algorithm}

\subsection*{Parallelism on Node Change Computations}
To leverage the advantage of modern multi-core computer systems, \lazyfox is capable of parallelizing the computations for the \emph{join} and \emph{leave}-actions by doing this in parallel for a queue of nodes. \autoref{lazyfox_flow} visualizes the parallel nature of a \lazyfox iteration. Every compute thread fetches a node from the node queue and computes the best actions independently. After the queue is emptied, \lazyfox gathers all the computed \emph{join}-actions and \emph{leave}-actions and applies all of them before re-filling the queue by the next set of nodes of the current iteration.

The core process for calculating node changes in \lazyfox is the same as in \fox (see \autoref{layzfox_iteration_algorithm}): The lines seven to 16 of \autoref{layzfox_iteration_algorithm} are the same as the lines three to eleven of \autoref{fox_iteration_algorithm}. The difference is the parallel computation of $queueSize$ many nodes in \lazyfox (lines one to six) and the bulk apply of node changes in lines 19 to 21.

The parallel computation introduces the first hyper-parameter of \lazyfox -- the queue size. The parallelization is what sets apart \lazyfox from \fox.
 
\algblockdefx[RUN]{Run}{EndRun}
  [1]{\textbf{run}~#1 }
  {\textbf{end run}}

\begin{algorithm}[H]
	\caption{\lazyfox} 
    \label{layzfox_iteration_algorithm}
	\begin{algorithmic}[1]
        \State nodeQueue = EmptyList()
	    \For{node x $\in$ Graph}
	        \State nodeQueue.insert(x)
	        \If {nodeQueue.size == queueSize \textbf{or} x is last node}
	            \For{node q $\in$ nodeQueue}
	                \Run{on separate thread}
	                    \State bestJoin = undefined
            			\For {community $c$ that does not contain $x$ and is a neighbor of $x$}
            				\State join = partition if $x$ joined $c$
            				\State update bestJoin if the join partition has a higher $\wcchat$ score
            			\EndFor
            	        \State bestLeave = undefined
            			\For {community $c$ that contains $x$}
            				\State leave = partition if $x$ left $c$
            				\State update bestLeave if the leave partition has a higher $\wcchat$ score
            			\EndFor
	                \EndRun
	            \EndFor
	            \State collect bestLeaves and bestJoins as nodeChanges from threads
                \State apply nodeChanges to the partitioning
                \State nodeQueue = EmptyList()
	        \EndIf
	    \EndFor
	\end{algorithmic}
\end{algorithm}

\subsection*{Stopping Criteria}
While computing and applying the \emph{join}-actions and \emph{leave}-actions, \lazyfox updates the \wcchat of the partition. At the end of each iteration, the current \wcchat is compared to the \wcchat of the partition an iteration earlier. If the total relative improvement of \wcchat is below a \wcchat-threshold -- a second hyperparameter -- \lazyfox stops and switches over to the post-processing phase. The first iteration of \lazyfox is compared to the \wcchat of the initial node clustering described in the  Preprocessing paragraph.

\subsection*{Post-processing}
As \lazyfox allows nodes to leave communities, communities can degenerate to having less than two members. This means that no triangle can be formed with that community anymore. Therefore, after each iteration, we remove such degenerated communities with less than two members.
In \lazyfox' final result, individual communities can not only overlap but also contain each other. The overlap can also lead to duplicated communities. Depending on the use-case, such duplicated or contained communities can be unwanted and have to be removed as part of the post-processing. While we provide an API to include custom post-processing of the final result, \lazyfox does not post-process the communities other than removing degenerated ones.

\subsection*{Evaluation Criteria}
\lazyfox improves the runtime of the community detection via the \fox approach. To evaluate the quality of the results from \lazyfox we utilize
overlapping normalized mutual information (ONMI) distance\footnote{OverlappingNMIDistance, \url{https://networkit.github.io/dev-docs/python_api/community.html?highlight=overlappingnmi\#networkit.community.OverlappingNMIDistance}, retrieved \date{15.09.2021}} and an appropriate F1 score\footnote{CoverF1Similarity, \url{https://networkit.github.io/dev-docs/python_api/community.html?highlight=coverf1similarity\#networkit.community.CoverF1Similarity}, retrieved \date{15.09.2021}}, two quality measurements that are widely adopted in the field of overlapping community detection.

\paragraph{Overlapping NMI Distance}
We use a recent variation of the common Normalized Mutual Information to enable its use for overlapping clusterings: Overlapping Normalized Mutual Information (ONMI) \citep{ONMI}.
The core idea of the ONMI score is to compare each combination of a community from the \lazyfox output with each community in the ground truth. We then select for each output community the closest community in the ground truth leveraging \emph{lack of information} as our distance function. Then the sum over minimal distances is averaged across all output clusters yielding one score, the ONMI distance, ranging from 0 (best) to 1 (worst). We employ its implementation via the OverlappingNMIDistance function in Networkit \citep{staudt2015NetworKit}.

\paragraph{F1-Score for Overlapping Communities}
To enable the use of the common F1 score for community partitions we use the approach proposed by \citep{Epasto2017f1overlapping} and implemented as CoverF1Similarity function in Networkit \citep{staudt2015NetworKit}. Every community is matched to its best ground truth community, and the resulting F1 scores are then averaged to form the final F1-Score, ranging from 0 (worst) to 1 (best). A single overlapping community $C'$ out of the detected partition $P'$ is compared with the regular F1-Score to all ground truth communities $C$ out of the ground truth partition $P$. The maximum F1 score is then chosen as the F1 score for $C'$:

\[F1-Score(P', P) = \frac{1}{|P'|} \sum_{C' \in P'} \max_{C \in P} F_1(C', C)\]

The F1 score for comparing a detected community $C'$ and a ground truth community $C$ is defined via the precision $p(C', C) = \frac{|C \cap C'|}{|C'|}$ and recall $r(C', C) = \frac{|C \cap C'|}{|C|}$:

\[F_1(C', C) = 2 \cdot \frac{p(C', C) \cdot r(C', C)}{p(C', C) + r(C', C)}\]
\section*{Datasets}
\label{sec:datasets}
To validate both our \fox and \lazyfox implementation we collected a set of well-known ``by bond" \citep{Ren2007} networks and ran the algorithms on them to either benchmark their performance or test whether our implementations are capable of dealing with the respective input sizes. 

All of our datasets were obtained from the SNAP-datasets \cite{snapnets} collection (see \autoref{dataset_size_overview} for an overview). The networks are provided in an edge list format.
In the following subsections, we present the specific datasets that we used with both their semantic and structural properties. 

\begin{table}
\centering
\begin{tabular}{|c|c|c|c|c|}
\hline
Dataset& Short Identifier & Node Count& Edge Count& Average Degree\\
\hline
Eu-core & eu   & 1,005 & 25,571 & 25.4 \\
DBLP    & dblp    & 317,080 & 1,049,866 & 3.3\\
LiveJournal & lj & 3,997,962 & 34,681,189 & 8.7\\
Friendster  & friendster & 65,608,366 & 1,806,067,135 & 27.5\\
\hline
\end{tabular}
\caption{Dataset Size Overview}
\label{dataset_size_overview}
\end{table}

\subsection*{Eu-core}
Eu-core\snapdownloaded{email-Eu-core} \citep{snapnets} is our first and smallest benchmarking dataset. It consists of anonymized email data between members of a large European research institution. An edge exists between two members if at least one email has been sent between these members. The Eu-core graph consists of 1005 nodes and 25,571 edges. The network is very dense with a diameter of length $7$ and a 90-percentile effective diameter of $2.9$. This can be expected from any cooperation, as people working with each other are likely to communicate. 

\subsection*{DBLP}
The Digital Bibliography \& Library Project (DBLP) is a computer science publication database provided by DBLP Organisation. The Library consists of more than 5.4 million publications from different journals and conference articles.
The DBLP dataset\snapdownloaded{bigdata/communities/com-dblp.ungraph} \citep{snapnets}, the second of our benchmarking datasets, is the co-authorship network calculated for these publications. Each of the 1,049,866 edges in this dataset represents co-authorship on a publication between two of the 317,080 authors.

\subsection*{LiveJournal}
The LiveJournal dataset is derived from the LiveJournal social network. In this social network, users can assign mutual friendship and join user-defined groups. For the dataset\snapdownloaded{/bigdata/communities/com-lj.ungraph} \citep{snapnets} the 3,997,962 nodes represent users and the 34,681,189 edges are formed from the friendship relations between these nodes.

\subsection*{Friendster}
Like the LiveJournal dataset, the Friendster dataset\snapdownloaded{/bigdata/communities/com-friendster.ungraph} \citep{snapnets} is also derived from the social network bearing the same name. But the Friendster dataset is an order of magnitude larger bringing in 
65,608,366 nodes and 1,806,067,135 edges and 8.7GB of compressed edges. While this scale makes it unattractive for benchmarking \fox and \lazyfox, it serves us as a feasibility test for the algorithm and implementation on large scale.

\section*{Results}
The implementation of the \lazyfox algorithm enables us to evaluate the runtime improvements and the impact on overlapping community quality of a multithreaded approach. Note that running \lazyfox with a queue size of one is equivalent to the original \fox algorithm.

During the evaluation, the queue size was always equal to the thread count. The results of this section were computed with the thread counts in $\{1, 2, 4, 8, 16, 32, 64, 128, 256\}$ and a fixed \wcchat threshold of 0.01. The computations were made on an HPE XL225n Gen10 machine with 512 GB RAM, and two AMD EPYC 7742 with 64 cores.

\subsection*{Runtime Improvements}
The main goal of any multithreaded approach is to improve the runtime. \lazyfox enables the parallel calculation of node changes, speeding up this step. The execution of these changes, however, has to be done sequentially, which is the main non-parallelizable part of the algorithm. The runtime measurements were taken without saving the results of the computation to disk to remove the very volatile I/O operation bottleneck from the benchmarks. The preprocessing steps described in the methods section are part of the measurement.

\begin{figure}[htb]
\centering
\includegraphics[width=0.8\linewidth]{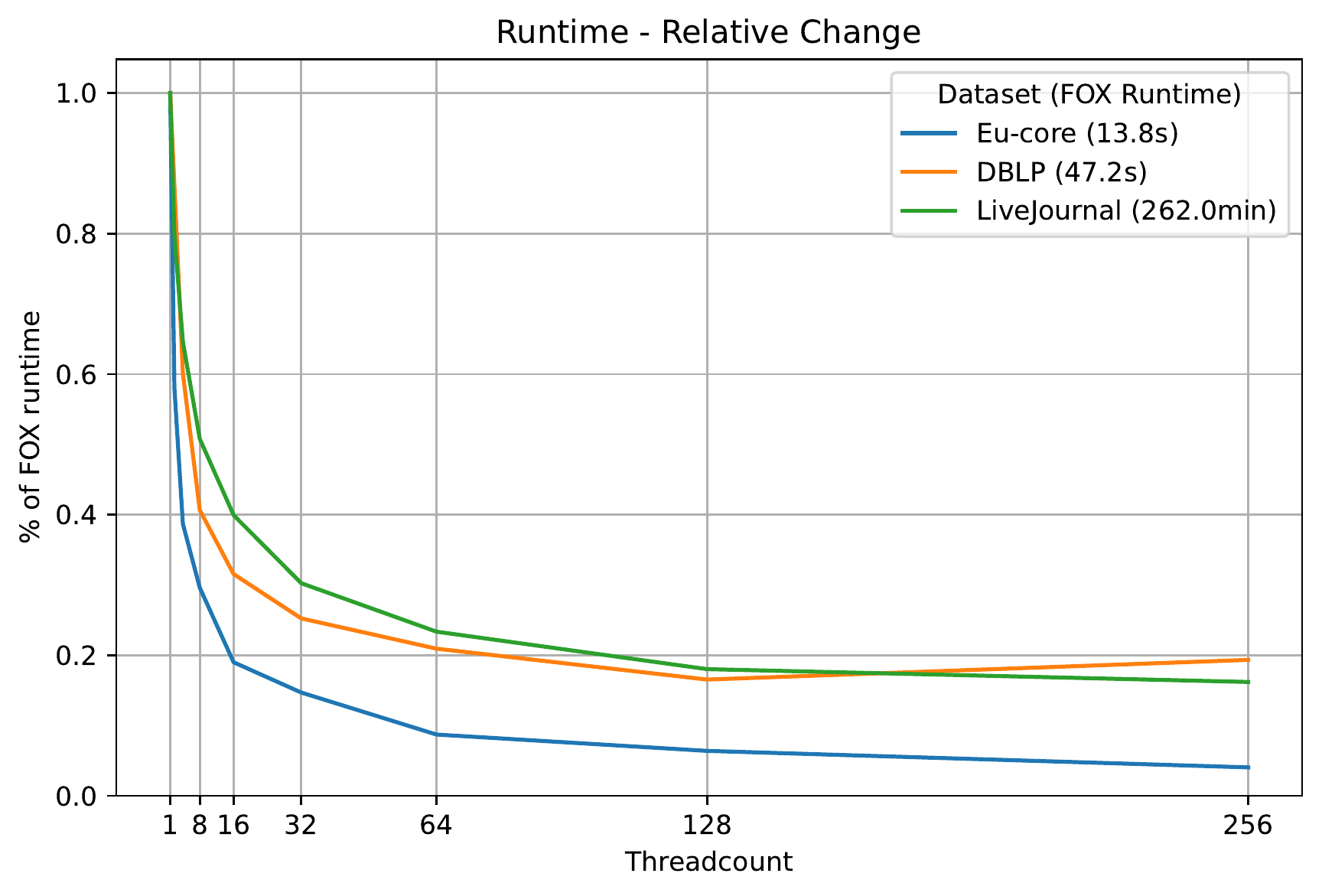}
\qquad
\begin{tabular}{|c||c|c|c|}
\hline
Threadcount& Eu-core & DBLP & LiveJournal\\
\hline

1 & 1.0 (13.8s)    & 1.0 (47.2s)    & 1.0 (262.0min)  \\
2 & 0.58    & 0.87   & 0.80  \\
4 & 0.39    & 0.60   & 0.65  \\
8 & 0.30    & 0.41   & 0.51  \\
16 & 0.19   & 0.32   & 0.40  \\
32 & 0.15   & 0.25   & 0.30  \\
64 & 0.09   & 0.21    & 0.23  \\
128 & 0.06  & 0.17    & 0.18  \\
256 & 0.04  & 0.19    & 0.16  \\
\hline
\end{tabular}
\caption{Runtime Comparison}
\label{runtimeComparision}
\small
The runtimes of \lazyfox on three different datasets EU-core, DBLP, and LiveJournal and different threadcounts relative to the non-parallelized runtime (threadcount $1$, \fox algorithm).

\end{figure}

We find significant improvements of runtime with the multithreaded approach on the Eu-core, DBLP and LiveJournal datasets (see \autoref{runtimeComparision}). Even the use of two threads instead of one reduces the original runtime of the \fox algorithm by about 20\%. While extra threads improve the runtime, the improvement rate does not stay the same.

The relative runtime improvements in the DBLP and the LiveJournal datasets are roughly the same, despite the fact that the LiveJournal dataset has orders of magnitude more nodes than the DBLP dataset. The original runtime on the Eu-core dataset (Eu) is reduced massively. With 256 threads \lazyfox takes just 4\% of the original \fox runtime. However, this extreme improvement is an exception to the overall results, due to the small scale of the Eu-core dataset ($\sim 1000$ nodes spread on $256$ threads).
Overall, we observe consistent runtime improvements also for larger scale networks, making the improvements independent of dataset size.

We also see that the runtime of \lazyfox on the DBLP dataset gets worse switching from 128 to 256 threads (see \autoref{runtimeComparision}). It is not caused by an additional iteration that had to be made by \lazyfox due to imprecision, as all DBLP runs take exactly seven iterations to converge. Instead, the increase in runtime appears before the first iteration runs, in the run preparations, such as initializing the threads. Thus, the most likely cause for the increased runtime is that the 128 additional thread initializations going from 128 threads in total to 256 threads in total do not yield enough runtime improvement to account for their own initialization time. 
Taken together, we see a thread count of 64 or 128 as optimal, being a trade-off between runtime improvement and resource usage for our examined networks.

\subsection*{Quality Impact}
\label{quality_impact}
The multithreaded approach of \lazyfox does alter the computations. Therefore, the results of \lazyfox on the same dataset differ for different threadcounts. However, we find these changes to be insignificant for sufficiently large datasets. The following sections compare quality measurements and analysis results of the singlethreaded approach to the multithreaded approaches.

\paragraph{Ground Truth Metrics}
To verify that the difference in community results of \lazyfox and \fox are negligible, we compare the results with each other (see \autoref{f1_and_nmi_compare}). We use two common metrics to measure overlapping community similarities, F1-Score and ONMI distance (see Methods). As the computational differences between \fox and \lazyfox increase with a higher degree of parallelism, we compare \lazyfox with increasing threadcounts.

We find that on sufficiently large datasets, \lazyfox results in communities extremely similar to the \fox communities (F1-Score of .99, ONMI distance of $\le .01$). However, we find that for the small Eu-core dataset, a high degree of parallelism does impact the result. While \lazyfox with a threadcount of $128$ creates results with .99 F1-Score and .05345 ONMI distance, the increase to $256$ threads impacts the final result, reducing the F1-Score to .85 and increasing the ONMI distance to 0.32.
These results suggest that smaller datasets such as Eu-core are more prone to changes caused by parallelization.

\begin{table}
\centering
\begin{tabular}{|c||c|c|c|c|c|c|}
\hline
Threadcount& \multicolumn{2}{|c|}{Eu-core}& \multicolumn{2}{|c|}{DBLP} & \multicolumn{2}{|c|}{LiveJournal}\\
\quad& F1 & ONMI-D & F1 & ONMI-D & F1 & ONMI-D\\
\hline
1   &1.0    &0.0 	    &1.0 	&0.0 	    &1.0 	&0.0\\
2   &0.99   &0.02386 	&0.99 	&0.00005 	&0.99 	&0.00348\\
4   &0.99   &0.03123 	&0.99 	&0.00007 	&0.99 	&0.00564\\
8   &0.99   &0.03005 	&0.99 	&0.00008 	&0.99 	&0.00684\\
16  &0.99   &0.03316 	&0.99	&0.00014 	&0.99 	&0.00756\\
32  &0.99   &0.05028 	&0.99 	&0.00017 	&0.99 	&0.00799\\
64  &0.99   &0.04884 	&0.99 	&0.00027 	&0.99 	&0.00837\\
128 &0.99   &0.05345 	&0.99 	&0.00034 	&0.99 	&0.00874\\
256 &0.85   &0.31553 	&0.99 	&0.00055 	&0.99 	&0.00950\\
\hline
\end{tabular}
\caption{Comparison of \lazyfox Results to \fox Results}
\label{f1_and_nmi_compare}
\small
Note that \lazyfox with a threadcount of $1$ is \fox. F1: F1-Score,  ONMI-D: ONMI distance, both calculated with the networkit library \citep{staudt2015NetworKit}.
\end{table}

\paragraph{Community Analysis}
We also perform a structural analysis of the results obtained by \lazyfox. In the best case, the properties measured (such as average community size) should stay the same, regardless of threadcount. Similar properties mean that all threadcounts can be used to draw the same conclusions over community structures within a specific dataset, making the degree of parallelism used in \lazyfox a hyper-parameter only relevant for run-time improvement, not for result quality.

The most obvious property of community detection is the number of detected communities. The threadcount used has no impact on this number, and for all threadcounts we detected 213 communities for the Eu-core, 76,513 communities for the DBLP and 920,534 communities for the LiveJournal dataset.

Looking at the size of the communities, the maximum and minimum sizes for DBLP and LiveJournal communitites do not change dependent on threadcount. The biggest community in DBLP has 115 members, LiveJournal 574. In the Eu-core dataset the threadcount changes the original maximum size of 135 members by maximum of plus two. However, the run with 256 threads is an exception, as the maximum size here is decreased by 39. This is another indication that a threadcount close to the node count of the dataset (256 is about one fourth of the Eu-core node count) is not optimal.

Finally, we can investigate the overlap between communities. This is the average number of communities a node belongs to. Again, this value does not change for DBLP or LiveJournal over different threadcounts, staying at 1.19 for DBLP and 1.9 for LiveJournal. The Eu-core dataset's overlap changes with different threadsizes, starting with an overlap of 5.44 for threadcount equal to one. For a threadcount up to four this stays relatively stable with changes of less than 0.1. However, higher threadcounts change the overlap quite drastically, up to +0.3. Threadcount 256 is the exception here again, changing the overlap from the original 5.44 to 4.22, which is one community less per node.

From these results we conclude that on sufficiently large datasets, the threadcount does not impact the obtained community structure. The changes only occur with the Eu-core dataset, and get worse with higher threadcounts. This is caused by the small node count in Eu-core. The nodecount therefore can be used as an indicator to determine the most favorable degree of parallelization. As the Eu-core results already change at two threads instead of one, we would suggest that the node count should at least be 500 times the threadcount to avoid these complications.

\subsection*{Cluster Analysis}
While it is worthwhile to investigate the impact of the threadcount to the various metrics, these metrics can also be used to analyze the semantics of the datasets.

\lazyfox optimizes triangle count. This means nodes end up in the same group if they are strongly connected with each other.
If we speak of people, this means they are socially close. In all our datasets the nodes represent people, connected by bond, such as common work (Eu-core, DBLP) or other social interaction (LiveJournal, Friendster). 

\subsubsection*{Eu-core}
We find 213 communities with an average size of 11. This means that within the research institution, there are about 213 teams with an average team size of 11 people. These teams can already be disbanded, as the data contains e-mails of a long timespan. The high overlap of 5.44 indicates that a single person works on average in 5.44 teams, however, because teams can already be disbanded this is not true for any point in time. It means that an employee works on average in 5.44 teams before leaving the institution or settling for a final research team. This could mean that there are only 5 or 6 team changes before reaching a leading position in a team or leaving the company, capping off the team changes.

\subsubsection*{DBLP}
We find 76,513 communities in the DBLP network with an average size of 4.9 members per community. Semantically these communities represent research groups. Scientists are the nodes, and edges are their co-authorship. Strong connections mean that any author of the group has published at least once with most of the other authors. The average size is reasonable for research groups, and indicates that larger research groups have sub-groups who rarely co-author with each other. Also, there is low overlap between groups, about 1.2, which means each author has on average 1.2 research groups. While one would assume a researcher stays true to his or her field, the nodes with an overlap over 1 are likely to be more senior members of the research community, having been part of multiple research groups, or researchers who changed their fields.

\subsubsection*{LiveJournal}
Another dataset we looked at is LiveJournal, a platform and social network for blogging. Users can befriend each other and join common interest groups. In this dataset we find that the average node has a degree of 17 and an overlap of 1.9. This means that each user has on average 17 friends and belongs to about two friendship groups. We assume these represent real life groups, as we think a mutual friendship circle of average 8 people (community size mean) is unlikely to form from social network interactions alone.
However, this is an assumption and verifying that would need sensitive data of LiveJournal user accounts to the people behind them and their real life friendships, which we do not have.

\subsection*{Large Scale Analysis -- Friendster}
\begin{figure}
\centering
\includegraphics[width=0.7\linewidth]{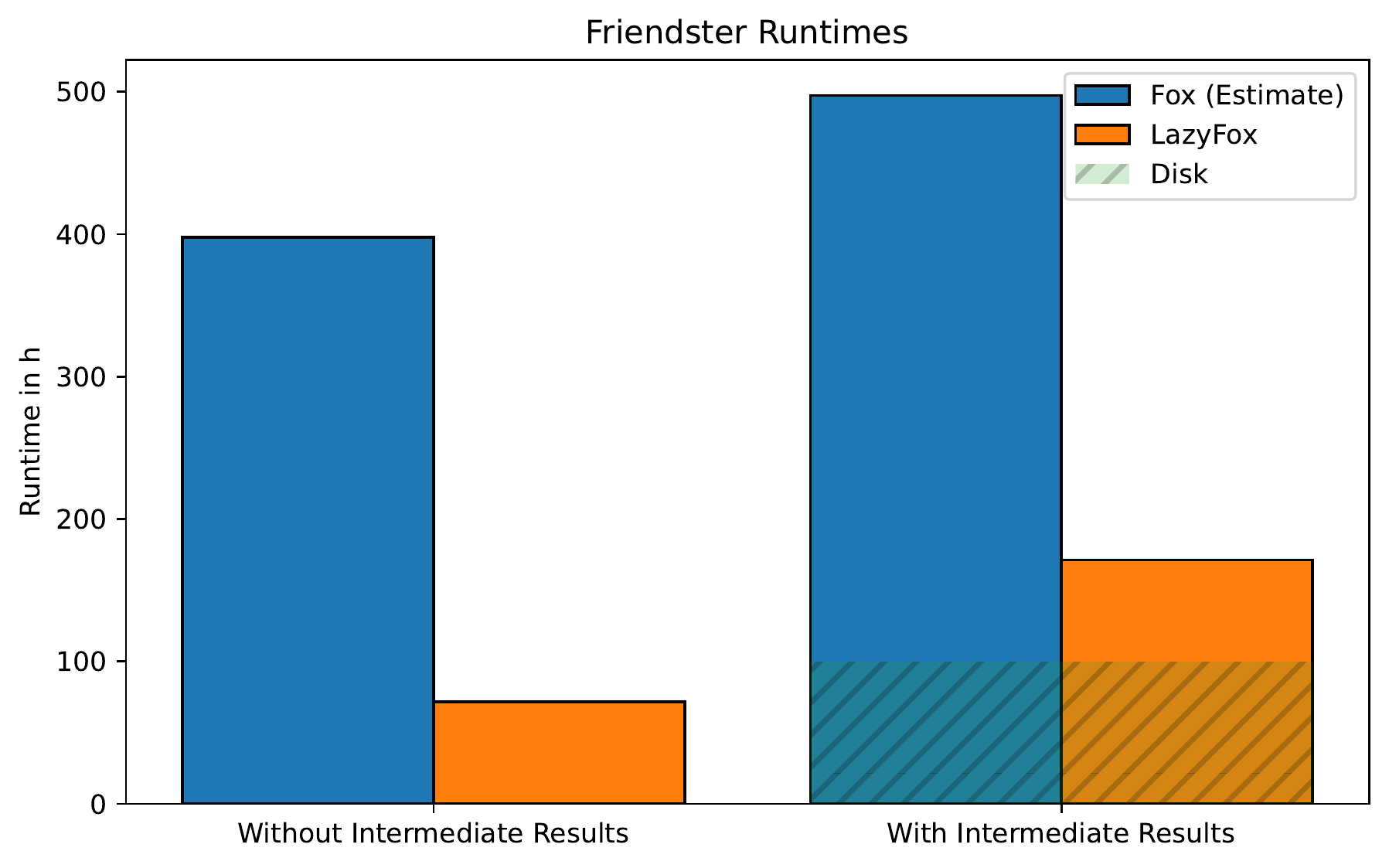}

\caption{Friendster Runtime Comparison}
\label{friendsterRuntimeComparision}
\small
The runtime of \lazyfox on the Friendster dataset with 256 threads compared to estimates of \fox runtime. Estimates made assuming a speedup comparable to the DBLP and LiveJournal datasets. \lazyfox' parallel computations make Friendster community analysis feasible.
\end{figure}

Finally, we would like to present the results of our computations on the Friendster dataset. The enormous scale of this dataset makes an analysis with \fox not feasible in reasonable time. However, due to the runtime improvements achieved with \lazyfox this analysis is made possible. We ran \lazyfox with 256 threads on the Friendster dataset twice, once  with and once without saving the intermediate results to disk. For both runs we use the measured \lazyfox runtime to estimate the \fox runtime \autoref{friendsterRuntimeComparision}. We do this by assuming a speedup ratio similar to the speedup ratios of the DBLP and LiveJournal datasets ($0.18$), as we found that the speedup is independent of scale at datasets of sufficient size.

The disk operations of saving the computation results of such a large dataset as Friendster are time consuming: After each iteration about 1.5 GB of clustering results are written to disk. \lazyfox converged after eight iterations, running about 171.3 hours ($\sim$ 7.1 days) with and 71.6 hours ($\sim$ 3.0 days) without saving to disk. It is save to say that the I/O operations and their preparations take up most of the runtime. This has to be kept in mind when working with large datasets.

\autoref{friendsterRuntimeComparision} illustrates why \lazyfox enables community analysis of large scale networks. A runtime of 71.6 hours ($\sim$ 3.0 days) is far more feasible than a runtime of almost 400 hours ($\sim$ 16.7 days).

After the 8th iteration, \lazyfox results in 13,861,732 detected communities, with an average size of 9 nodes per community. However, the largest community has 4,721 members. This indicates that Friendster was mostly used for real friend groups where a size of 9 people is reasonable, but it also contains closely connected communities of great size. We find that about 92\% of all communities have 2 to 19 members. Communities over 100 members make up only 0.003\% of all communities. We assume that these large communities are career networks, institutions or non-human user networks (bots), where a high interconnectedness between members is beneficial to the members, even if they do not know each other personally.
We also find that a node belongs on average to 1.92 communities, meaning each user has about 2 groups of friends.

We find \lazyfox highly viable to run analyses on such large datasets, despite the runtime of over a week. However, computing metrics such as F1 which compare the results to the graph itself would be challenging due to the sheer size of the graph.

\subsection*{\wcchat Threshold Impact}
Another hyper-parameter apart from the threadcount for the \lazyfox algorithm is the iteration termination criterion, the \wcchat threshold. \lyu propose a threshold of 0.01 of relative \wcchat change, meaning that if an iteration decreases the global \wcchat by less than 1\%, the algorithm stops. This threshold was determined by experiments measuring community size and density (compare \cite{fox}, Table 7), as a lower threshold did not yield significant change in these properties. 

\begin{figure}[hbt]
    \centering
    \includegraphics[width=0.8\linewidth]{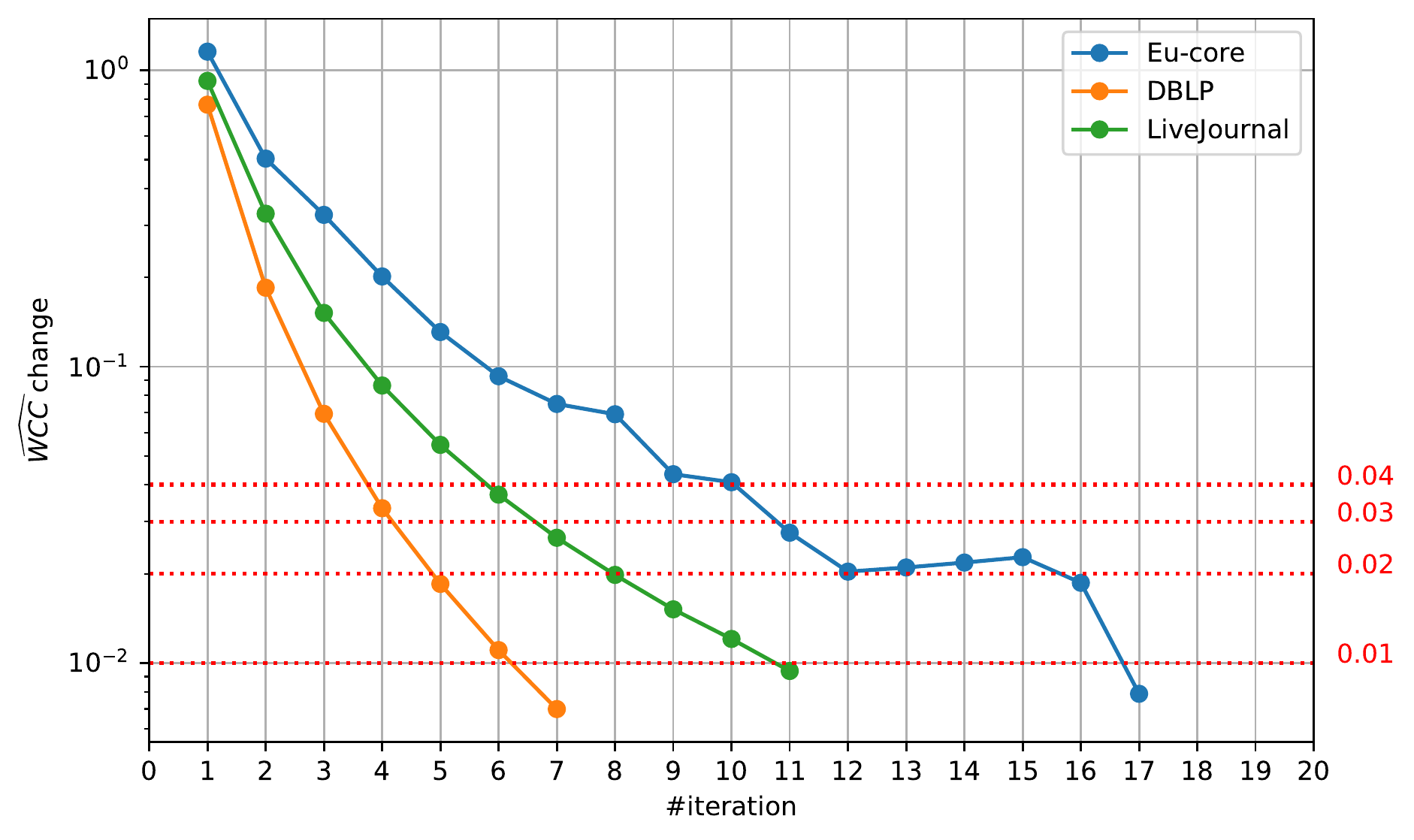}
    \caption{Relative \wcchat Change Development}
    \label{fig:relative_wcc_development}
    \small
    The relative change of \wcchat at different iterations on different datasets, EU-core, DBLP, and LiveJournal. By default and also in \cite{fox}, computations stop if the change drops below 0.01. Other possible thresholds (0.02, 0.03, 0.04) are visualized.
\end{figure}

However, we find that the threshold's impact on the algorithm's runtime is quite extraordinary. As \autoref{fig:relative_wcc_development} shows, a slightly higher threshold can save multiple iterations. This has high impact on runtime, as the iterations get slower over time. We find that even the threshold of 0.02 instead of 0.01 yields great runtime improvements for runs on a single thread.
For the Eu-core dataset, it saves five iterations, which is equivalent to 39\% of the total runtime. For DBLP, we save two, for LiveJournal three iterations, which is equivalent to 33\% and 43\% of the total runtime, respectively.

These runtime improvements are due to an earlier termination and therefore can still be seen in runs with a higher threadcount. Running \lazyfox with 256 threads on the LiveJournal dataset with a \wcchat threshold of 0.02 still saves 39\% of the runtime, compared to the same run with a \wcchat threshold of 0.01. However, the effectiveness can be dependent on the dataset scale, as the savings for the Eu-core and DBLP datasets at this high threadcount are significantly smaller (3\% and 10\%, respectively).

\begin{figure}
    \centering
    \includegraphics[width=\linewidth]{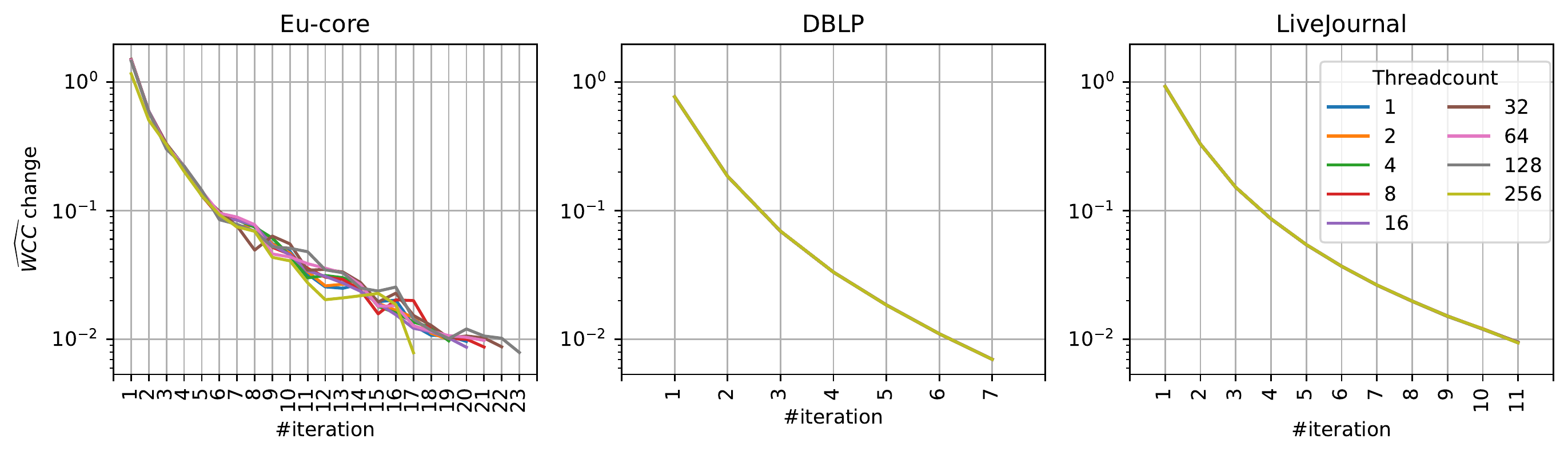}
    \caption{\wcchat at different threadcounts}
    \label{fig:wcc_over_time_over_threads}
    \small
    \wcchat development at different threadcounts at different iterations on different datasets, EU-core, DBLP, and LiveJournal. Threadcount does not influence the \wcchat development at sufficiently large datasets. Note that in DBLP and LiveJournal datasets the values for each threadcount are exactly the same.
\end{figure}

It is important to note that these decreased savings at higher threadcounts are due to the overall decreased total runtime and not due to a different \wcchat development over time (see  \autoref{fig:wcc_over_time_over_threads}). While a higher threadcount can lead to different results and therefore change the \wcchat scores, this is not the case at large scale datasets. As the community analysis results are unaffected by higher threadcounts (see 'Quality Impact' section), so is the \wcchat. The Eu-core dataset here again is an outlier due to the threadcount being close to its nodecount.

\section*{Discussion}
Our proposed parallelization of the algorithm slightly differs in its logic from the original version \fox \citep{fox}: \lazyfox computes first all node changes for a certain batch of nodes (in parallel) and then executes them altogether to generate a new community structure. In contrast, in \fox, computed changes for a single node are applied immediately. However, \lazyfox still generates highly similar results to \fox and therefore enables detecting meaningful overlapping communities. These results go along the same lines as the parallelization of e.g.\ the BigClam algorithm \citep{Liu2018SpeedingUB}. They show once more that many analysis approaches can be parallelized even if the resulting algorithms are not strictly equivalent to their sequential counterparts.

With the runtime improvement provided by the parallelization \lazyfox brings overlapping community detection into new areas of application and allows for the analyses of significantly larger and more complex datasets. It reduces the computation time for networks with millions of nodes and billions of edges from weeks to days. For example, \lazyfox could not only be used to find community structures in huge social media networks such as twitter, but also to find complexes or pathways in intricately connected multi-layered molecular network \citep{liu2020MultilayerBio}, or querying global-scale, finely resolved climate \citep{steinhaeuser2011climatesystem}, or ecological networks \citep{bohan2017biomonitoring}. \lazyfox is very flexible, the algorithm can be straightforwardly adapted to take directed or weighted graphs into account. We would like to address these extensions in the future, thereby opening even further application areas for overlapping community detection.

 \lazyfox relies on nodes iteratively joining and leaving communities. Therefore, it could generate empty, disconnected, or communities fully contained in other communities. We provide an additional post-processing function to eliminate those types of undesired results. However, note that it can be computationally expensive to perform this post-processing on large graphs, or graphs with many detected communities. Sometimes, communities that are fully contained in other communities are desirable and occur in the ground truth. In these cases, e.g. the citation network in SNAP \citep{Yang2011CommunityMetricEval}, they should not be removed. The post-processing needed is therefore dictated by the research domain which is why detailed analyses on post-processing is not part of our work.
 
 We showed that the value of the \wcchat  threshold hyperparameter is decisive for the runtime of the algorithm. This is expected as it constitutes the termination criterion of this iterative algorithm, and it is comparable to criteria applied to convergence assessment for other graph-based analyses, such as the PageRank algorithm in NetworKit \citep{staudt2015NetworKit}. Alternatively, limiting the number of performed steps is another solution for termination of iterative algorithms that has been applied previously, e.g.\ in node-centric PageRank computation in iPregel \citep{Capelli2019iPregel}. An investigation of the asymptotic behavior, i.e.\ whether the \wcchat score of \lazyfox converges, might be interesting, but we have not experienced problems in the application cases we analyzed. As heuristic solution, we suggest to monitor the change in \wcchat over the course of the computation to discover problematic cases.
 
The queue size as second hyperparameter of \lazyfox was set to the available threadcount in our experiments. In principle, \lazyfox allows to choose both values independently from one another. However, increasing the threadcount over the queue size does not yield any runtime improvements as the queue size controls the maximum degree of parallelization. Additional threads will then actually slow down the computation as their allocation takes time.

Moreover, the larger the queue size, the more different \lazyfox becomes to \fox in theory, and we cannot guarantee that the community detection results remain stable. Overall, we find that the optimal queue size and thread count should scale with the size of the dataset, but unless very small graphs are analysed, up to 250 cores can be exploited without strong impact on the results.

\section*{Conclusion}
We provide \lazyfox, an open-source implementation of an efficient, parallelized algorithm for overlapping community detection. Our implementation and analysis is another positive example for the idea that many graph analysis algorithms can be parallelized, even if not being strictly equivalent to their sequential version.
The results show that the in-parallel computed optimization of the \wcchat metric yields extremely similar results to the sequential computation. This allows the leverage of modern hardware to significantly decrease runtime, enabling the detection of community structures of very large graphs, without loosing community quality. The results on the impact of the \wcchat threshold allows an informed decision on the runtime - \wcchat score trade-off.
With our improvements, we make overlapping community detection achievable for very large graphs with at least tens of million nodes and a few billion edges in a reasonable amount of time. Thus, we enable this type of graph analysis for novel and complex application domains that have not been previously explored with respect to overlapping community detection.

\section*{Acknowledgments}
We would like to thank Martin Taraz for his excellent support and discussions. KB furthermore acknowledges funding from an Add-on Fellowship for Interdisciplinary Life Sciences of the Joachim Herz Stiftung.

\bibliography{lib}

\begin{thebibliography}{}

\bibitem[Ahn et~al., 2010]{Ahn2010MultiscaleComplex}
Ahn, Y.-Y., Bagrow, J., and Lehmann, S. (2010).
\newblock Link communities reveal multiscale complexity in networks.
\newblock {\em Nature}, 466:761--4.

\bibitem[Airoldi et~al., 2007]{airoldi2007mixed}
Airoldi, E.~M., Blei, D.~M., Fienberg, S.~E., and Xing, E.~P. (2007).
\newblock Mixed membership stochastic blockmodels.

\bibitem[Barabasi et~al., 2011]{Barabasi2011NetworkMedicine}
Barabasi, A.-L., Gulbahce, N., and Loscalzo, J. (2011).
\newblock Network medicine: A network-based approach to human disease.
\newblock {\em Nature reviews. Genetics}, 12:56--68.

\bibitem[Barabasi and Oltvai, 2004]{Barabasi2004NetworkBiology}
Barabasi, A.-L. and Oltvai, Z. (2004).
\newblock Network biology: Understanding the cell's functional organization.
\newblock {\em Nature reviews. Genetics}, 5:101--13.

\bibitem[Basuchowdhuri et~al., 2014]{Basuchowdhuri2014ProductRecommendation}
Basuchowdhuri, P., Shekhawat, M.~K., and Saha, S.~K. (2014).
\newblock Analysis of product purchase patterns in a co-purchase network.
\newblock In {\em 2014 Fourth International Conference of Emerging Applications
  of Information Technology}, pages 355--360.

\bibitem[Boccaletti et~al., 2006]{Boccaletti2006ComplexNetworks}
Boccaletti, S., Latora, V., Moreno, Y., and Hwang, D.-U. (2006).
\newblock Complex networks: Structure and dynamics.
\newblock {\em Physics Reports}, 424:175--308.

\bibitem[Bohan et~al., 2017]{bohan2017biomonitoring}
Bohan, D., Vacher, C., Tamaddoni-Nezhad, A., Raybould, A., Dumbrell, A., and
  Woodward, G. (2017).
\newblock Next-generation global biomonitoring: Large-scale, automated
  reconstruction of ecological networks.
\newblock {\em Trends in Ecology \& Evolution}, 32:477–487.

\bibitem[Bu et~al., 2017]{Bu2017GLEAM}
Bu, Z., Cao, J., Li, H., Gao, G., and Tao, H. (2017).
\newblock Gleam: a graph clustering framework based on potential game
  optimization for large-scale social networks.
\newblock {\em Knowledge and Information Systems}, 55:741--770.

\bibitem[Capelli et~al., 2019]{Capelli2019iPregel}
Capelli, L.~A., Hu, Z., Zakian, T.~A., Brown, N., and Bull, J.~M. (2019).
\newblock ipregel: Vertex-centric programmability vs memory efficiency and
  performance, why choose?
\newblock {\em Parallel Computing}, 86:45--56.

\bibitem[Chakraborty, 2015]{Chakraborty2015LeveragingDC}
Chakraborty, T. (2015).
\newblock Leveraging disjoint communities for detecting overlapping community
  structure.
\newblock {\em ArXiv}, abs/1504.06608.

\bibitem[Danon et~al., 2005]{Danon2005ComparingCommunityI}
Danon, L., Duch, J., Diaz-Guilera, A., and Arenas, A. (2005).
\newblock Comparing community structure identification.
\newblock {\em Journal of Statistical Mechanics: Theory and Experiment}, 2005.

\bibitem[Epasto et~al., 2017]{Epasto2017f1overlapping}
Epasto, A., Lattanzi, S., and Leme, R. (2017).
\newblock Ego-splitting framework: from non-overlapping to overlapping
  clusters.
\newblock pages 145--154.

\bibitem[Evans and Lambiotte, 2009]{evans2009LinkPartitions}
Evans, T. and Lambiotte, R. (2009).
\newblock Line graphs, link partitions and overlapping communities.
\newblock {\em Physical review. E, Statistical, nonlinear, and soft matter
  physics}, 80:016105.

\bibitem[Fortunato, 2010]{Fortunato2010}
Fortunato, S. (2010).
\newblock Community detection in graphs.
\newblock {\em Physics Reports}, 486(3-5):75--174.

\bibitem[Gavin et~al., 2006]{Gavin2006YeastModularity}
Gavin, A.-C., Aloy, P., Grandi, P., Krause, R., Boesche, M., Marzioch, M., Rau,
  C., Jensen, L., Bastuck, S., Dümpelfeld, B., Edelmann, A., Heurtier, M.-A.,
  Hoffman, V., Hoefert, C., Klein, K., Hudak, M., Michon, A.-M., Schelder, M.,
  Schirle, M., and Superti-Furga, G. (2006).
\newblock Proteome survey reveals modularity of the yeast cell machinery.
\newblock {\em Nature}, 440:631--6.

\bibitem[Gopalan and Blei, 2013]{Gopalan14534}
Gopalan, P.~K. and Blei, D.~M. (2013).
\newblock Efficient discovery of overlapping communities in massive networks.
\newblock {\em Proceedings of the National Academy of Sciences},
  110(36):14534--14539.

\bibitem[Gregory, 2009]{Gregory2009FindingOC}
Gregory, S. (2009).
\newblock Finding overlapping communities in networks by label propagation.
\newblock {\em ArXiv}, abs/0910.5516.

\bibitem[Guimerà and Amaral, 2005]{Guimera2005FunctMetabolic}
Guimerà, R. and Amaral, L. (2005).
\newblock Functional cartography of complex metabolic networks.
\newblock {\em Nature}, 23:22--231.

\bibitem[Guimerà et~al., 2005]{Guimera2005AirTransportNetworks}
Guimerà, R., Mossa, S., Turtschi, A., and Amaral, L. (2005).
\newblock The worldwide air transportation network: Anomalous centrality,
  community structure, and cities' global roles.
\newblock {\em Proceedings of the National Academy of Sciences of the United
  States of America}, 102:7794--9.

\bibitem[Hofman and Wiggins, 2008]{hofman2008bayesModularity}
Hofman, J.~M. and Wiggins, C.~H. (2008).
\newblock Bayesian approach to network modularity.
\newblock {\em Physical Review Letters}, 100(25).

\bibitem[Kelley et~al., 2012]{Kelley2012SocialCommunities}
Kelley, S., Goldberg, M.~K., Magdon-Ismail, M., Mertsalov, K., and Wallace, A.
  (2012).
\newblock Defining and discovering communities in social networks.

\bibitem[Lancichinetti and Fortunato,
  2009]{Lancichinetti2009ComparativeComDetection}
Lancichinetti, A. and Fortunato, S. (2009).
\newblock Community detection algorithms: A comparative analysis.
\newblock {\em Physical review. E, Statistical, nonlinear, and soft matter
  physics}, 80:056117.

\bibitem[Lancichinetti et~al., 2009]{Lancichinetti2009DetectingTO}
Lancichinetti, A., Fortunato, S., and Kert{\'e}sz, J. (2009).
\newblock Detecting the overlapping and hierarchical community structure in
  complex networks.
\newblock {\em New Journal of Physics}, 11:033015.

\bibitem[Lancichinetti et~al., 2011]{Lancichinetti2011OSLOM}
Lancichinetti, A., Radicchi, F., Ramasco, J.~J., and Fortunato, S. (2011).
\newblock Finding statistically significant communities in networks.
\newblock {\em PloS one}, 6:e18961.

\bibitem[Lee et~al., 2010]{Lee2010DetectingHO}
Lee, C., Reid, F., McDaid, A.~F., and Hurley, N.~J. (2010).
\newblock Detecting highly overlapping community structure by greedy clique
  expansion.
\newblock In {\em KDD 2010}.

\bibitem[Leskovec and Krevl, 2014]{snapnets}
Leskovec, J. and Krevl, A. (2014).
\newblock {SNAP Datasets}: {Stanford} large network dataset collection,
  http://snap.stanford.edu/data.

\bibitem[Liu and Chamberlain, 2018]{Liu2018SpeedingUB}
Liu, C. H.~B. and Chamberlain, B.~P. (2018).
\newblock Speeding up bigclam implementation on snap.
\newblock {\em 2018 Imperial College Computing Student Workshop (ICCSW 2018)}.

\bibitem[Liu et~al., 2020]{liu2020MultilayerBio}
Liu, X., Maiorino, E., Halu, A., Glass, K., Prasad~B, R., Loscalzo, J., Gao,
  J., and Sharma, A. (2020).
\newblock Robustness and lethality in multilayer biological molecular networks.
\newblock {\em Nature Communications}, 11.

\bibitem[Lyu et~al., 2020]{fox}
Lyu, T., Bing, L., Zhang, Z., and Zhang, Y. (2020).
\newblock Fox: Fast overlapping community detection algorithm in big weighted
  networks.
\newblock {\em ACM Transactions on Social Computing}, 3:1--23.

\bibitem[Ma et~al., 2019]{Ma2019PPIdrugsdiscovery}
Ma, J., Wang, J., Soltan~Ghoraie, L., Men, X., Haibe-Kains, B., and Dai, P.
  (2019).
\newblock A comparative study of cluster detection algorithms in
  protein–protein interaction for drug target discovery and drug repurposing.
\newblock {\em Frontiers in Pharmacology}, 10.

\bibitem[Mcauley and Leskovec, 2014]{Mcauley2014SocialCircles}
Mcauley, J. and Leskovec, J. (2014).
\newblock Discovering social circles in ego networks.
\newblock {\em ACM Trans. Knowl. Discov. Data}, 8(1).

\bibitem[McDaid et~al., 2011]{ONMI}
McDaid, A., Greene, D., and Hurley, N. (2011).
\newblock Normalized mutual information to evaluate overlapping community
  finding algorithms.
\newblock {\em CoRR}.

\bibitem[Newman, 2006]{Newman2006Modularity}
Newman, M. (2006).
\newblock Modularity and community structure in networks.
\newblock {\em Proceedings of the National Academy of Sciences of the United
  States of America}, 103:8577--82.

\bibitem[Palla et~al., 2005]{palla2005cfinder}
Palla, G., Derényi, I., Farkas, I., and Vicsek, T. (2005).
\newblock Uncovering the overlapping community structure of complex networks in
  nature and society.
\newblock {\em Nature}, 435:814--818.

\bibitem[Prat-P\'{e}rez et~al., 2012]{wcc}
Prat-P\'{e}rez, A., Dominguez-Sal, D., Brunat, J.~M., and Larriba-Pey, J.-L.
  (2012).
\newblock Shaping communities out of triangles.
\newblock In {\em Proceedings of the 21st ACM International Conference on
  Information and Knowledge Management}, CIKM '12, page 1677–1681, New York,
  NY, USA. Association for Computing Machinery.

\bibitem[Prat-P\'{e}rez et~al., 2014]{PratPerezSCD}
Prat-P\'{e}rez, A., Dominguez-Sal, D., and Larriba-Pey, J.-L. (2014).
\newblock High quality, scalable and parallel community detection for large
  real graphs.
\newblock In {\em Proceedings of the 23rd International Conference on World
  Wide Web}, WWW '14, page 225–236, New York, NY, USA. Association for
  Computing Machinery.

\bibitem[Psorakis et~al., 2011]{Psorakis2011OverlappingCD}
Psorakis, I., Roberts, S., Ebden, M., and Sheldon, B.~C. (2011).
\newblock Overlapping community detection using bayesian non-negative matrix
  factorization.
\newblock {\em Physical review. E, Statistical, nonlinear, and soft matter
  physics}, 83 6 Pt 2:066114.

\bibitem[Raghavan et~al., 2007]{Raghavan2007LabelPropagation}
Raghavan, N., Albert, R., and Kumara, S. (2007).
\newblock Near linear time algorithm to detect community structures in
  large-scale networks.
\newblock {\em Physical review. E, Statistical, nonlinear, and soft matter
  physics}, 76:036106.

\bibitem[Regan and Barabasi, 2003]{Regan2003HierachyComplexNetworks}
Regan, E. and Barabasi, A.-L. (2003).
\newblock Hierarchical organization in complex networks.
\newblock {\em Physical Review E}, 67.

\bibitem[Regan et~al., 2002]{Regan2002HierarchicalMetabolic}
Regan, E., Somera, A., Mongru, D., and Oltvai, Z. (2002).
\newblock Hierarchical organization of modularity in metabolic networks.
\newblock {\em Science}, 297.

\bibitem[Reid et~al., 2011]{Reid2011PartitioningBC}
Reid, F., McDaid, A.~F., and Hurley, N.~J. (2011).
\newblock Partitioning breaks communities.
\newblock In {\em ASONAM}.

\bibitem[Ren et~al., 2007]{Ren2007}
Ren, Y., Kraut, R., and Kiesler, S. (2007).
\newblock Applying common identity and bond theory to design of online
  communities.
\newblock {\em Organization Studies}, 28(3):377--408.

\bibitem[Saltz et~al., 2015]{Saltz2015DistributedCD}
Saltz, M., Prat-P{\'e}rez, A., and Dominguez-Sal, D. (2015).
\newblock Distributed community detection with the wcc metric.
\newblock {\em Proceedings of the 24th International Conference on World Wide
  Web}.

\bibitem[Shi et~al., 2013]{Shi2013ALC}
Shi, C., Cai, Y., Fu, D., Dong, Y., and Wu, B. (2013).
\newblock A link clustering based overlapping community detection algorithm.
\newblock {\em Data Knowl. Eng.}, 87:394--404.

\bibitem[Song et~al., 2015]{Song2015}
Song, Y., Bressan, S., and Dobbie, G. (2015).
\newblock {\em Fast Disjoint and Overlapping Community Detection}, pages
  153--179.
\newblock Springer Berlin Heidelberg, Berlin, Heidelberg.

\bibitem[Staudt et~al., 2015]{staudt2015NetworKit}
Staudt, C., Sazonovs, A., and Meyerhenke, H. (2015).
\newblock Networkit: A tool suite for large-scale complex network analysis.

\bibitem[Steinhaeuser et~al., 2011]{steinhaeuser2011climatesystem}
Steinhaeuser, K., Ganguly, A., and Chawla, N. (2011).
\newblock Multivariate and multiscale dependence in the global climate system
  revealed through complex networks.
\newblock {\em Climate Dynamics}, doi:10.1007/s00382-011-1135-9.

\bibitem[Wang and Wong, 1987]{Wang1987StochasticBF}
Wang, Y.~J. and Wong, G. Y.~C. (1987).
\newblock Stochastic blockmodels for directed graphs.
\newblock {\em Journal of the American Statistical Association}, 82:8--19.

\bibitem[Watts and Strogatz, 1998]{Watts1998}
Watts, D.~J. and Strogatz, S.~H. (1998).
\newblock Collective dynamics of `small-world' networks.
\newblock {\em Nature}, 393(6684):440--442.

\bibitem[Xie and Szymanski, 2012]{Xie2012OverlapLabelProp}
Xie, J. and Szymanski, B. (2012).
\newblock Towards linear time overlapping community detection in social
  networks.
\newblock volume abs/1202.2465.

\bibitem[Yang and Leskovec, 2012]{Yang2011CommunityMetricEval}
Yang, J. and Leskovec, J. (2012).
\newblock Defining and evaluating network communities based on ground-truth.
\newblock {\em Knowledge and Information Systems}, 42.

\bibitem[Yang and Leskovec, 2013]{Yang2013OverlappingCD}
Yang, J. and Leskovec, J. (2013).
\newblock Overlapping community detection at scale: a nonnegative matrix
  factorization approach.
\newblock In {\em WSDM '13}.

\end{thebibliography}

\end{document}